\documentclass[10pt]{article}
\usepackage[utf8]{inputenc}
\usepackage[T1]{fontenc}

\usepackage{amsmath}
\usepackage{graphicx}
\usepackage{float}
\usepackage{multicol} 
\usepackage{url}
\usepackage[a4paper,left=2cm,right=2cm,top=2cm,bottom=4cm,bindingoffset=5mm]{geometry}

\title{The concerted emergence of well-known spatial and temporal ecological patterns in an evolutionary food web model in space}

\author{
  Michaela Hamm\thanks{\texttt{mhamm@fkp.tu-darmstadt.de}}\,  and Barbara Drossel \\
Institut f\"ur Festk\"orperphysik, TU Darmstadt \\  
  Hochschulstr. 6 \\
  64289 Darmstadt
}

\begin{document}
\maketitle

\begin{abstract}
Ecological systems show a variety of 
characteristic patterns of biodiversity in space and time. 
It is a challenge for theory to find models that can reproduce and explain the observed patterns. 
Since the advent of island biogeography these models revolve around speciation, dispersal, and extinction, but they usually neglect trophic structure. 
Here, we propose and study a spatially extended evolutionary food web model that allows us to study large spatial systems with several trophic layers. Our computer simulations show that the model gives rise simultaneously to several biodiversity patterns in space and time,  from species abundance distributions to the waxing and waning of geographic ranges.
We find that trophic position in the network plays a crucial role when it comes to the time evolution of range sizes, because the trophic context restricts the occurrence and survival of species especially on higher trophic levels.
\end{abstract}


\section{Introduction}
Patterns in space and time have sparked the interest of ecologists for decades, if not centuries.
In a pattern lies the quiet promise of some underlying, fundamental ``truth''.
A general mechanism that, if deciphered, will naturally lead to an explanation of the pattern.

Patterns in ecology emerge on a variety of spatial and temporal scales\cite{Levin1992}.
The most studied patterns in space are species abundance distributions (SADs), range size distributions (RSDs), similarity decay, and species area relationships (SARs).
SADs show the classical feature of few abundant and many rare species, probably following a log-normal or similar function\cite{McGill2007}.
RSDs show that range sizes of species are normally small\cite{Brown1996}, in agreement with the shape of the SAD\cite{Gaston1996, Gaston2003}.
Related to small ranges is the decay of the similarity of communities.
With geographical distance similarity between communities usually decays exponentially\cite{Nekola1999, Soininen2007}.
Overlapping species distributions lead to SARs, where species number increases with area as a power-law or a similar function\cite{Tjoerve2003} with exponents between 0.2 - 0.4 \cite{Drakare2006}. 

Temporal patterns are more difficult to obtain, as they require the aquisition of data from the fossil record.
Notable are species lifetime distributions that appear to follow a power-law with an exponent around -1.6\cite{Newman1999}, and species range evolution curves that show a ``hat-shaped''\cite{Zliobaite2017a} pattern of waxing and waning in the geographic range of a species over its lifetime\cite{Liow2007, Foote2007}.
An investigation of the causal relation between area and lifetime concluded that the influence is mutual, as species that live longer can spread further, and  species that are spread further have a smaller risk of going extinct by local fluctuations\cite{Foote2007}.
Recently the relative importance of the impact of the environment and of competition was studied finding that the tip of the curve is more influenced by competition whereas the rising and falling flanks are more influenced by environmental conditions\cite{Zliobaite2017a}.

Another ``pattern'', or rather a rule, of ecology was formulated as the law of constant extinction\cite{Valen1973}, which says that the extinction probability of a species does not depend on its age because species co-evolve in ever lasting competition and thus can never gain advantage over competitors (``Red Queen hypothesis'').
But when taking range expansion into account, the Red Queen hypothesis must be rejected or at least modified because there is a clear indication that species occupying larger ranges have smaller chances of extinction\cite{Finnegan2008}.

An important challenge for theoretical ecology consists in finding models that unify these various spatial and temporal patterns and explain them by common underlying mechanisms. 
A model should explain several patterns at once \cite{McGill2010}. 
The key processes that shape biodiversity patterns have long been identified to be speciation, extinction, and dispersal\cite{Rosenzweig1995}, whilst a unified underlying model is still missing.
This debate is reflected in a large number of co-existing theories as summarized by McGill\cite{McGill2010}.
McGill compares six different models and extracts three ingredients shared by all models, which are needed for a theory on biodiversity patterns to successfully reproduce diversity patterns \textit{in space} (SAR, distance decay of similarity)\cite{McGill2010}.
Those are hollow curve SAD (implying that one empirical pattern is explicitly presupposed in order to obtain other patterns), clumping of populations in space, and a lack of spatial correlation of populations.
Those features emerged in a simple assembly model near regional diversity equilibrium that also reproduced spatial ecological patterns, like SAR and range size distributions\cite{OSullivan2019}.
May and coauthors \cite{May2016} showed that these three features are not sufficient to reproduce patterns in empirical data of tropical forests. 
They found that a model that was fitted to show the features proposed by McGill did well in reproducing the SAR but failed for distance decay and co-occurrence patterns.
The authors concluded that the three features are not sufficient and that the requirement of spatial uncorrelatedness might be violated in real systems due to habitat preferences\cite{May2016}.

The theories mentioned so far revolve around spatial patterns in diversity.
Temporal patterns are often analysed separately, although there is a connection of range and lifetime\cite{Willis1922, Gaston2000}, (but see \cite{Rogge2019, Bertuzzo2011} for examples that incorporate both spatial and temporal scales). 
In our opinion, however, one should aim at unifying spatial and temporal patterns and thus obtain a deeper understanding of mechanisms that shape biodiversity patterns.

Another aspect that is often neglected in studies of biodiversity patterns is the food web character of local communities. 
Especially for temporal patterns obtaining fossil food web data is nearly impossible, considering how much effort needs to be put into the reconstruction of a single food web from a single lagerstätte\cite{Dunne2014}.
Models often dodge the challenge of incorporating food web structure by using competitive communities\cite{McPeek2007, OSullivan2019} or neutral models which are valid for only one trophic layer\cite{Durrett1996, Rosindell2007}.

One notable example is the model of Rogge and coauthors that analysed SARs and lifetime distributions in an evolutionary food web model in space\cite{Rogge2019}.
However, as their model did not incorporate biomasses of species they could not analyse SADs. Neither  did they provide information on spatial co-occurrence or distance decay of similarity.
Evolutionary food web models that incorporate explicit population dynamics run into computational problems (time constrains) when large spatial and temporal scales shall be simulated.
As a consequence, previous evolutionary food web models only include a small number of habitats, see for example\cite{Allhoff2015a, Loeuille2008, Bolchoun2017}.

The model we propose can serve as a valuable tool in exploring the connections between temporal and spatial macroecological patterns for complex trophic networks, as it can be used to model evolutionary time scales for large areas.
The model produces a large variety of empirical well studied patterns, like SAD, RSD, SAR, lifetime distributions, and range expansion evolution at once.
We further utilise the model to analyse how species range evolution depends on trophic position in the food web.
The nature of a food web entails that each species is embedded in a biotic environment spanned by the other species and at the same time competing with species on the same trophic level for resources.
So this approach offers the chance to study competition and environmental effects at the same time.
We find that basal species that are mainly governed by competition often show a classical waxing and waning pattern whilst consumers that need to deal with competition and environment (need for suitable layer of basal species) show more variance in their range history and have smaller ranges.

\section{Model}
We use an evolutionary food web model that merges elements from the classical Webworld model \cite{Caldarelli1998} with elements from more recent body-mass structured food web models \cite{Loeuille2005,Allhoff2013,Allhoff2015,Rogge2019}.
The elements from the Webworld model allow us to assign biomasses to all populations without performing time-consuming population dynamics calculations. In this way, we can compute network configurations on many habitats, i.e., on large spatial scales, over evolutionary times. While the Webworld model characterises species by a vector of traits, our model is based on the body mass as the only trait, and on correlated secondary traits such as the preferred prey size. 
The body mass has a clear biological meaning and a high importance for almost all processes in biology, for example metabolism \cite{Brown2004}, locomotion \cite{Hirt2017} or prey choice \cite{Brose2010}, which makes it a reasonable choice for a master trait.

\begin{figure}[t]
\centering
    \includegraphics[width = 0.75\linewidth]{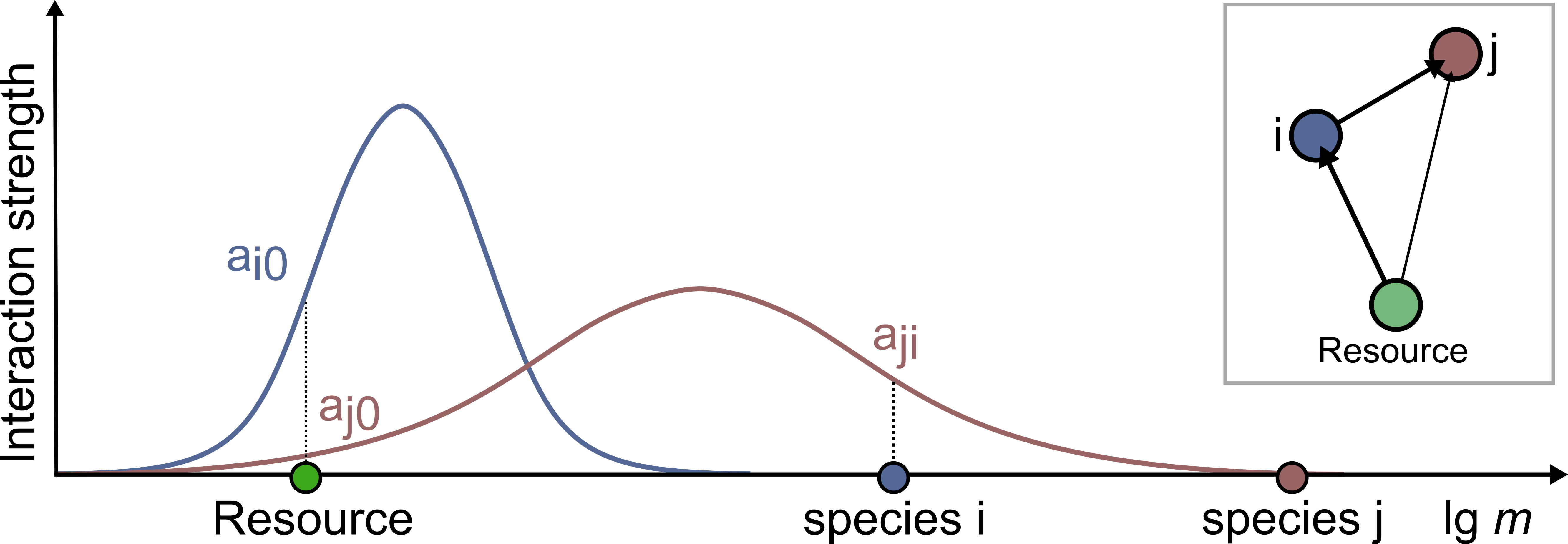}
    \caption{Illustration of the food web construction. Species are determined by their body mass and Gaussian feeding kernels. Any species inside this feeding kernel is possible prey for a species. The inset shows the resulting food web.}
    \label{fig: food web model}
\end{figure}

\paragraph{Traits and biomasses}
A species $i$ is specified by its body mass $m_i$, feeding center $f_i$, and feeding width $s_i$. The interaction strength $a_{ij}$ of a predator $i$ with prey $j$ is given by a Gaussian distribution with cutoff
\begin{equation}
    a_{ij}= e^{-\frac{(f_i-m_j)^2}{2s_i}} \theta(|f_i-m_j|/s_i)\, 
\end{equation}
similar to \cite{Allhoff2015},  see Fig. \ref{fig: food web model} for illustration.

As in the Webworld model, the actual feeding rate is obtained by taking competition between predators of the same prey into account.
We denote the set of all predators of a single prey $j$ as $P_j$.
The ``best'' predator (characterised by the highest feeding strength $a^{*}$) gets most of the prey. 
The proportion of prey obtained by each predator depends on its feeding strength difference $\Delta a_{i} = a^{*} - a_{ij}$ via
\begin{equation}
F_{ij} =  \exp \left( - \left( \frac{\Delta a_i}{\delta} \right)^2 \right).
\end{equation}

And after normalising this equation we obtain
\begin{equation}
\gamma_{ij} = \frac{F_{ij}}{\sum\limits_{k \in P_{j}} F_{k,j}}.
\end{equation}
The parameter $\delta$ allows a tuning of the competition strength. This expression differs from the Webworld model, where a linear function $\gamma_{ij} \sim 1 - \frac{\Delta a_i}{\delta}$ was used. 
This turned out to be a competition function that is not soft enough for a niche space with only one trait (body mass) and lead to mere food chains or to overcrowded basal layers with our feeding rules. So we replaced it by the more smooth Gaussian function, keeping the original idea of comparing predators with the best adapted predator.

Given the interaction parameters $\gamma_{ij}$, population sizes are calculated in the same way as in the Webworld model. The authors of that model proposed the following self-consistent relation for the  equilibrium population sizes $B(i)$:
\begin{equation}
B(i) = \gamma_{i, 0} R + \lambda \left( \sum_j \gamma_{ij} B(j) - \gamma_{ii} B(i) \right).\label{eq: population}
\end{equation}
The idea is that energy, which is limited and modelled by the resource parameter $R$, is passed from bottom to top throughout the network in a donor control fashion.
This means the biomass on the lower level completely determines how much biomass will be in the level above, i.e., there are no top-down effects.
The parameter $\lambda$ allows us to tune which fraction of biomass is transferred to predators.
Note that we use the term 'predator-prey interaction' even when dealing more generally with consumers and their food source, which may also be a herbivore-plant interaction.
Equation \eqref{eq: population} permits us also to specify a survival criterion: A species survives if its biomass density lies above the extinction threshold which we choose to be 1 as in the original Webworld publication.

\paragraph{Speciation and colonisation}
New species enter the system locally by variation of existing species.
In each time step a habitat is chosen randomly that will be the source of the event.
On this habitat a population is randomly chosen, that acts as a 'parent' for a 'child' species.
The body mass $m$ of the child species $m$ is chosen randomly from an interval around the parent's body mass, $m \in \left[q^{-1} m_{\text{parent}},\, q m_{\text{parent}} \right]$ (logarithmically scaled body masses), such that the child's biomass differs at most by a factor $q = 5$ from the parent's body mass.
The feeding center $f$ of the child is chosen from an interval below its body mass $f \in \left[m - 3,\, m - 1 \right]$.
The feeding range is random-uniformly chosen from the interval $\left[0.5, 1 \right]$.
Table \ref{tab:Parameters} summarises all numerical values used in the simulations.
The child is added to the habitat of the parent and is taken into consideration for the next calculation of the biomass densities via equation (\ref{eq: population}).

The other process that can bring new species into a habitat is dispersal. We use a square lattice of  habitats for our simulations. 
For a dispersal event, a species is randomly chosen and copied to an adjacent habitat.
We keep the average speciation rate per habitat constant at 1, thus taking it as the reference time scale.
Dispersal rate determines how many colonisation events occur on average per habitat  between two speciation events.
The dispersal rate is the most important control parameter of our model. 

After each speciation and dispersal event, equation \eqref{eq: population} is applied to the affected habitat in order to obtain the new population sizes. Over the course of time, the system self-organizes itself, with the resulting food webs and spatial species distribution being an emergent feature of the model. Since the model produces an ongoing turnover of species, it allows us to evaluate the spreading and withdrawal of species in time, in addition to species-area relationships, rank-abundance distributions, and lifetime distributions.   

\begin{table}[]
\centering
\caption{Overview of model parameters with corresponding values used in simulations. Variable $m$ in the feeding center definition refers to body mass of a species.}
\begin{tabular}{|l|c|c|}
\hline 
\rule[-1ex]{0pt}{2.5ex} \textbf{Parameter} & \textbf{Symbol} & \textbf{Numerical value}\\ 
\hline 
\rule[-1ex]{0pt}{2.5ex} Efficiency & $\lambda$ & 0.65  \\ 
\hline 
\rule[-1ex]{0pt}{2.5ex} Competition strength & $\delta$ & 0.25  \\ 
\hline 
\rule[-1ex]{0pt}{2.5ex} Resource per habitat & $R$ & 25  \\ 
\hline 
\rule[-1ex]{0pt}{2.5ex} Speciation width & $q$ & 5 \\ 
\hline 
\rule[-1ex]{0pt}{2.5ex} Feeding center interval & $f$ &  $\left[m - 3, m - 1 \right]$ \\ 
\hline 
\rule[-1ex]{0pt}{2.5ex} Feeding range interval & $\bar{r}$ & $\left[0.5, 1\right]$  \\ 
\hline 
\end{tabular} 
\label{tab:Parameters}
\end{table}

\section{Results}
\begin{figure}[t]
    \includegraphics[width = 1.0\linewidth]{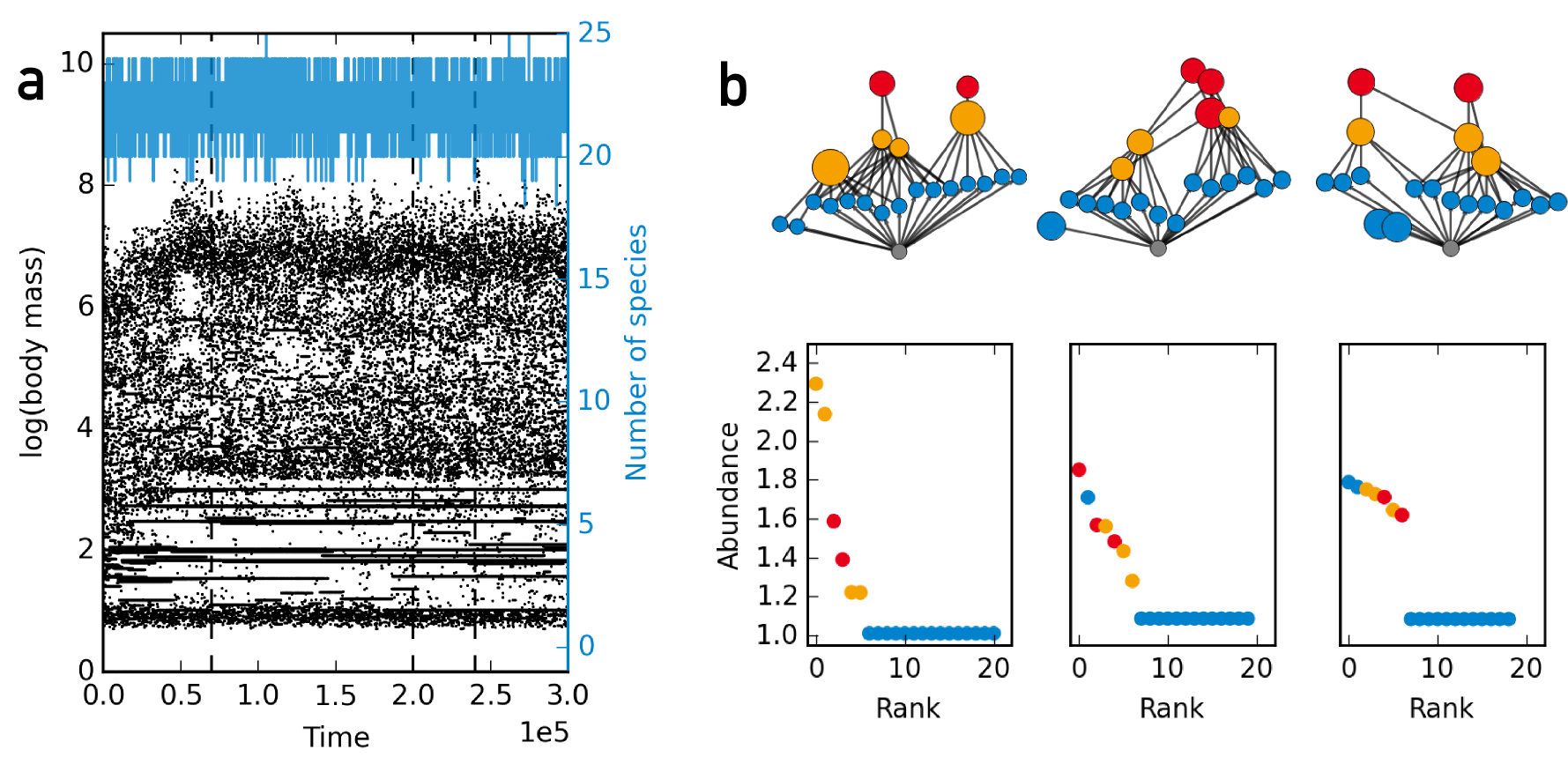}
    \caption{
    (\textbf{a}) Time series for one habitat out of a grid of 10$\times$10 habitats. At each time point the  body masses of all species present at this habitat are plotted. 
    The blue curve denotes species number in this habitat.
    (\textbf{b}) Food web structure and rank abundance curves for the three time points indicated in (a) by dashed vertical lines.
    Rank is determined by sorting species for their abundance. 
    This results in the most abundant species having the smallest rank. 
    Colors indicate trophic level (blue = basal, yellow = intermediate, red = top).
    } 
    \label{fig: time series example}
\end{figure}

\paragraph{Time Series}
Fig. \ref{fig: time series example}a shows a time series for a habitat that is part of a system with one hundred habitats.
At each point in time the body masses of all species in this habitat are plotted. Species enter the habitat either through speciation or dispersal and can persist only if their biomass according to equation \eqref{eq: population}  is larger than 1 in this habitat.
Species that persist longer in the habitat result in longer horizontal lines, while points indicate species that appear only for a short time. 
The number of species, shown in blue, fluctuates around 22, which is not far from the maximum species number of 25, determined by the combination of resource size and extinction threshold. 

Fig.~\ref{fig: time series example}b shows the food webs and the rank abundance curves for three points in time, which are marked by vertical dashed lines in panel a.
The shape of the rank abundance curve resembles the empirical well documented ``hollow curve'' shape \cite{McGill2007}: most species have small abundances just above the extinction threshold and a few species have large abundances. This is the ``most species are rare'' observation. 
The difference in abundances between ``rare'' and ``abundant'' species is smaller than in empirical systems. For the first example, the most abundant species has a biomass density that is 2.3 times larger than the least abundant species.
The data show also that basal species are less abundant than those on higher trophic levels in our model.
Higher trophic species have larger populations because they have less competitors for food than basal species, which are the most numerous.

We think that the difference to empirical findings in the range of abundance values is due to the fact that the trait space of our model is essentially one-dimensional, while real ecosystems allow for a large number of trait combinations and types of niches. 
Furthermore, if we would group species with the same predator and prey sets together as trophic species, the number of different basal species would be reduced and the biomass densities of these trophic species would increase. 

The most important feature of our model is that it, despite its simplicity, produces a continuous species turnover with several persisting trophic levels. The simulation shows no large extinction avalanches, nor does the dynamics freeze into a fixed species configuration. These features are crucial for exploring species-area distributions and the change of species rank and range in time. 

\begin{figure}[t]
\centering
\includegraphics[width = \linewidth]{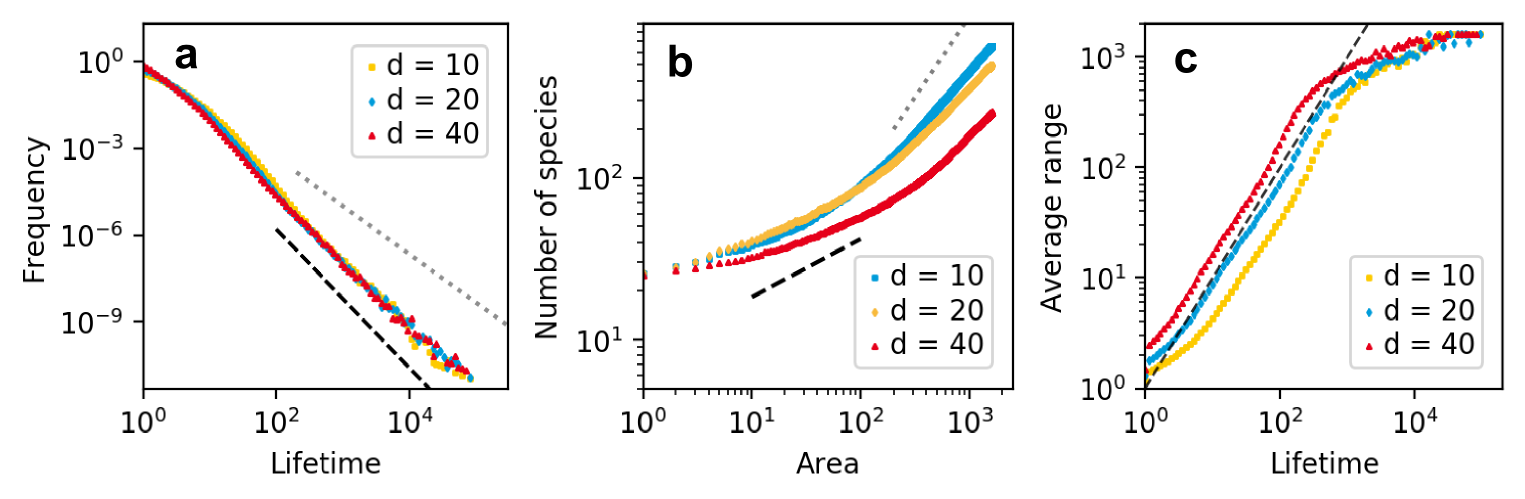}
    \caption{Patterns emerging in a square lattice of 1600 habitats for different dispersal rates $d$:
    (\textbf{a}) Lifetime distributions are broad and close to a power law with exponent $-2.4$ (black, dashed).
    (\textbf{b}) SAR reach empirically reasonable values (black dashed, slope of 0.36 \cite{Drakare2006}) for intermediate areas. For large areas curves bend toward a slope of 1 (grey, dotted). Dispersal rate decreases the slope of the SAR. 
    (\textbf{c}) Average area (number of occupied habitats over the lifetime of a species) increases with lifetime. Larger dispersal rate shifts the curve to larger areas.
    Dashed line has a slope of 1. 
    }
    \label{fig: SAR and LTD}
\end{figure}

\paragraph{Lifetimes and Areas}
Fig. \ref{fig: SAR and LTD} shows several relationships involving area and lifetimes for different dispersal rates on a grid with 1600 habitats simulated for $10^5$ speciation events per habitat. 
The lifetime of a species is the length of the time interval from its first appearance (which happens by a speciation process) until its global extinction from all habitats. 
The lifetime distributions (Fig. \ref{fig: SAR and LTD}a) are broad and fall off in a power-law like manner for longer lifetimes. 
These broad distributions manifested themselves already in Fig. \ref{fig: time series example}a in the vastly different lengths of the  horizontal lines. 
The slope of the curve is similar for all dispersal rates shown.
For visual guidance we plot a slope of $-2.4$ (black dashed) which is steeper than the slope of $-5/3$ (grey dotted) that was reported for a simpler spatial food web model\cite{Rogge2019}.

The corresponding species-area relationships are shown in Fig. \ref{fig: SAR and LTD}b.
For visual reference we include a slope of 0.36 (black dashed), which is an empirically reasonable average value \cite{Drakare2006}, and a slope of 1 (grey dotted), which is the slope for the limit that the considered area is much larger than the largest species range. 
With increasing area, all slopes must eventually approach this limit value. 
With increasing dispersal rate the slope decreases, and the upwards bend moves to larger areas. 
This is to be expected, as species ranges increase with increasing dispersal rate.
SARs are empirically well studied\cite{Drakare2006}.
As our model does not resolve the location of individuals within a habitat, our SARs show only the regional and the global scale, but not the local scale, where the slope should be again steeper  \cite{Rosindell2007}. 
The slopes observed for our model in the regional range have empirically reasonable numerical values.

Fig. \ref{fig: SAR and LTD}c shows the relation between lifetime and  the average range of a species, averaged over all species that fall in a small lifetime bin (width in log scale 0.055).
We measure ``range'' in number of occupied habitats, so the average range is the average number of habitats that a species occupied during its lifetime.
The dashed curve has a slope of 1.
We observe that species that live longer occupy larger ranges, closely following the slope of 1. This means that the number of habitats occupied by the species increases linearly with time.
This trend ends when the area approaches the system size of 1600 habitats.
Higher dispersal rate shifts the average range to higher values, as species disperse farther before a competitor replaces them.

\begin{figure}[t]
    \centering
    \includegraphics[width = 1.0\linewidth]{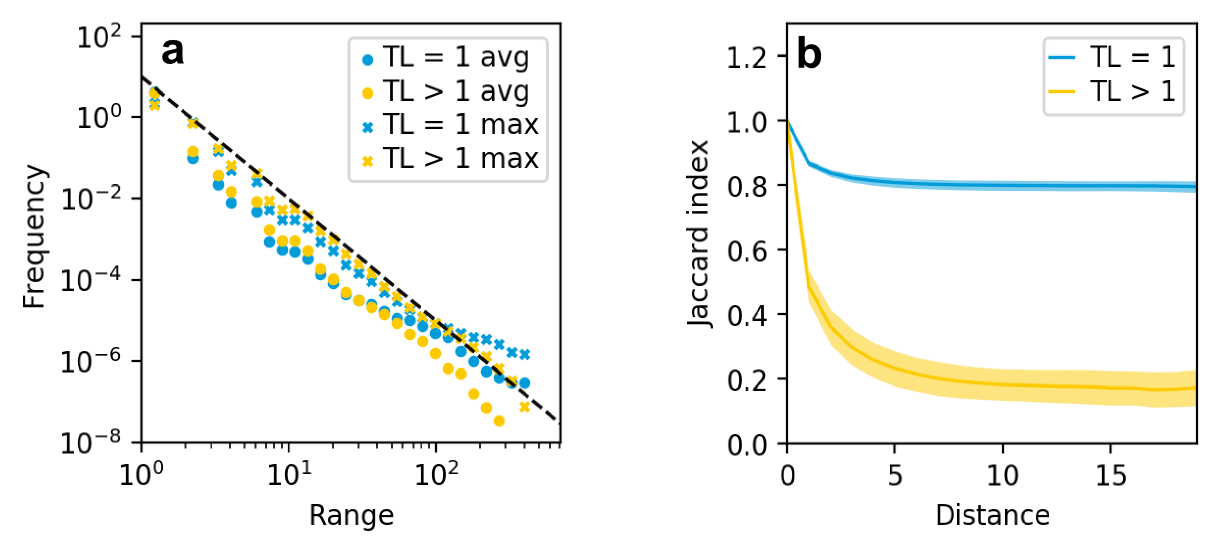}
    \caption{ Range distribution and similarity decay with distance in a system with 400 habitats and a dispersal rate of 10.
    (\textbf{a}) Distribution of maximum and average range for the basal layer (TL1) and all other species (> TL1). Average range refers to the average number of habitats a species occupied during its lifetime.
    Dashed line has a slope of -3, so most species have small ranges.
    (\textbf{b}) Food web similarity, measured by the Jaccard index, averaged over ten points during the same simulation as in (a), errorbars denote standard deviation from the mean. Similarity decreases with distance. The decrease is small for the basal layer, but steep for the higher trophic levels.
 }
    \label{fig:Range distribution}
\end{figure}

Fig. \ref{fig:Range distribution}a shows the distribution of the average and maximum range in a simulation with 400 habitats and a dispersal rate of 10.
The maximum range is the maximum number of habitats that a species occupied during its lifetime.
For orientation the black dashed line indicates a slope of -3. The data show that most species are only present on a few habitats and that basal species are on average farther distributed than non-basal species.
We even see a small fraction of basal species occupying all habitats of the system during most of their lifetime. 
These species have long lifetimes, as seen in Fig. \ref{fig: SAR and LTD}a.
Basal species have the advantage that they find resources everywhere in the system.
Species from higher trophic levels depend on the species configuration on the layers below, which makes it harder to conquer larger areas and have an extended lifetime.
Nevertheless, we see from the distribution of the maximum range that some species from higher trophic levels are present in all habitats at some point in time.

The range distribution does not give information on how the ranges of species look. 
Inhabited habitats could be clumped together or be patchily distributed across the grid.
The right panel of Fig. \ref{fig:Range distribution} shows the similarity of the local food webs as a function of distance between habitats.
We measure distance in minimum number of steps between two habitats and similarity as the Jaccard index of species composition between two habitats, i.e., the number of species shared by the habitats divided by total number of all species on both habitats.
Habitat similarity within the basal layer falls off over the first two habitats and then stabilises for larger distances at around 0.8. 
This is not unexpected, as we observed already that part of the basal species is present in all habitats during most if its lifetime, and that this lifetime is almost as long as the whole simulation time. 
The success of basal species is mainly determined by their own trait set (i.e. how well they are adapted to the resource) as this is the basis dividing the resource between consumers.

The habitat similarity on higher trophic levels shows the same shape as for the basal species, but decreases more smoothly over a larger distance and levels off at a similarity of 0.2.

Overall, food webs on neighbouring habitats are similar, which indicates that ranges of species that are not so widely distributed do not contain many holes.
This is in line with the shallow slope in the SAR for small areas: The species number increases only slowly with area if surrounding communities are similar.
For larger areas the composition of the food web changes and the slope gets steeper. 

\begin{figure}[t]
    \centering
    \includegraphics[width = 0.9\linewidth]{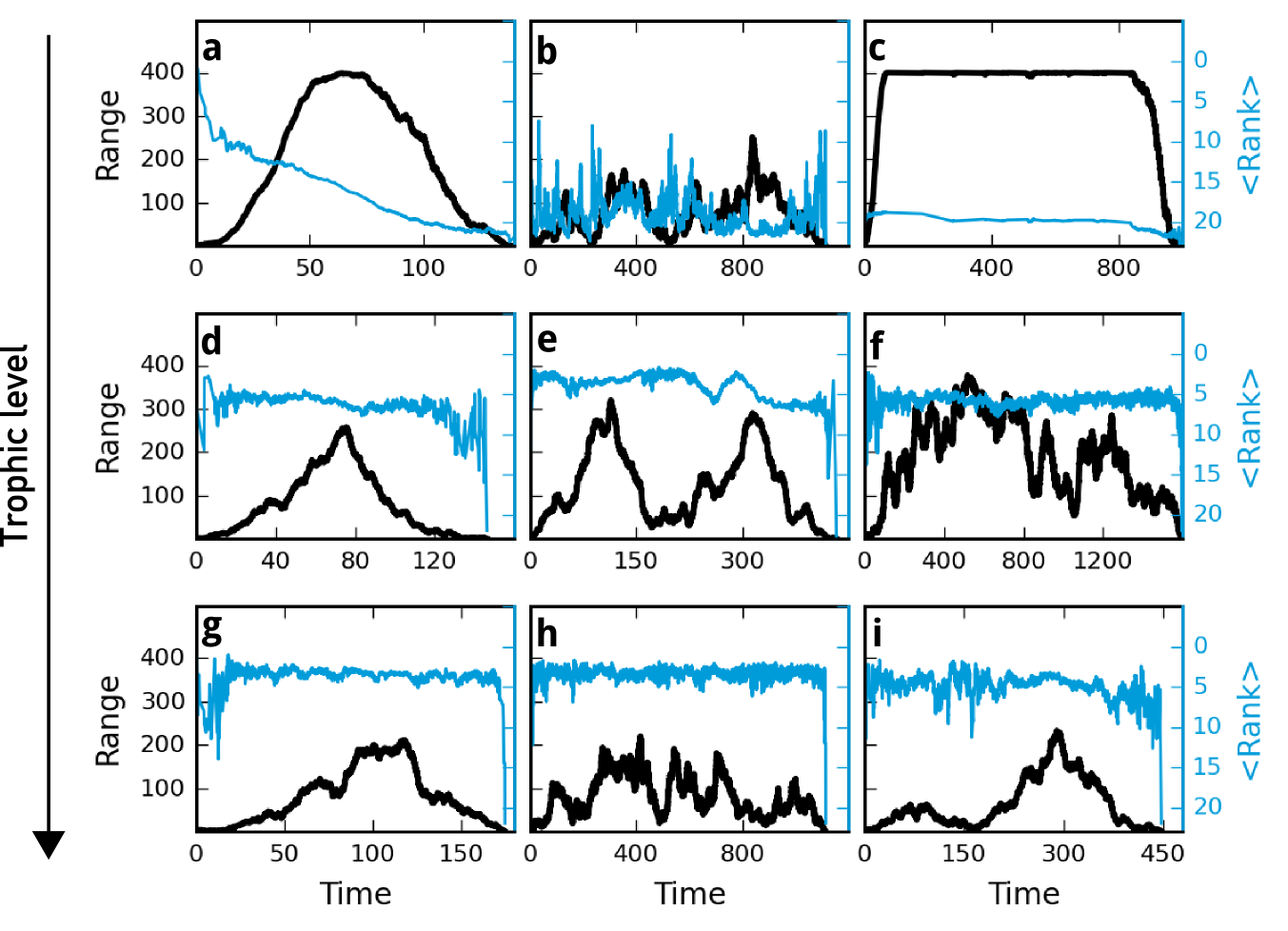}
    \caption{Collection of time evolution examples for range (black) and rank (blue) of exemplary species from different trophic levels. Simulation set up was 400 habitats and a dispersal rate of 10.
    Trophic level increases from top to bottom line ((a)-(c): trophic level 1, (d)-(f): trophic level 2, (g)-(i): trophic level $\geq$ 3). Range evolution shows a triangular shape in some cases, especially often in the basal layer, whilst other species have large fluctuations in their range and rank over time.}
    \label{fig:collection of hats}
\end{figure}

\paragraph{Geographic Ranges}
We now turn to the waxing and waning of the  geographic range over the lifetime of a species.
We focus on species that conquered at least half of the spatial grid at some point in their lifetime. 
Fig.\ref{fig:collection of hats} shows a collection of typical time series for geographic range (black) and rank (blue) of species from different trophic levels in a system with 400 habitats.
All trajectories share one feature: The geographic range starts to grow right after the emergence of the species.
We do not observe a species that resides for a long time in a small range and then suddenly starts to expand.
The shape of the range expansion curves can be grouped in two different patterns. (i) The ``hat'' pattern: A steady increase in range size up to a maximum range, then a steady decrease to extinction. 
We also count curves that show small peaks of in- and decrease on the back of a hat shape as hat-shaped.
This category also includes the cases where the species conquers the whole network and resides for some time on all habitats.
In a larger spatial system this species would keep increasing its range and thus show a hat curve.
(ii) The multiple peak pattern: Some range evolution curves show no distinct hat shape.
They look more like random walk patterns.
Noteworthy are cases in which species can recover from a loss of range and possibly reach their maximum range during a later peak.

The shape of the range expansion curves depends on trophic position. Basal species mostly show the hat pattern.
The lifetime and range distributions already indicated that some basal species live long and prosper, some can retain a viable population on all habitats for a very long time, for which we see an example in Fig. \ref{fig:collection of hats}c.
Intermediate species as well as top species show a large variety of hat and non-hat patterns.
Those species also show more frequently the described recovery from a very small range to an even higher range (bottom right panel).
We also observe that the higher the trophic level the smaller the maximum range (peak), which is in line with the range distributions (see Fig.~\ref{fig:Range distribution} above).
Regarding the rank evolution of species we observe a large variety of curves.
Basal species often show a constant or steadily decreasing rank over time.
A decrease of rank means that the average biomass density decreases. Such a steady decrease can be used as an early warning sign that indicates the chance of extinction of those species.
Intermediate species and top species show a less predictable rank evolution, consistent with the more diverse range evolution plots.
Large fluctuations in rank indicate that species have vastly different ranks in different habitats.
When species loose habitats where they are strongly present, the average rank changes considerably.
Here, the time evolution of the rank cannot be used as an early warning sign, as it does not follow a predictable pattern before extinction.

\begin{figure}[t]
    \centering
    \includegraphics[width = 0.6\linewidth]{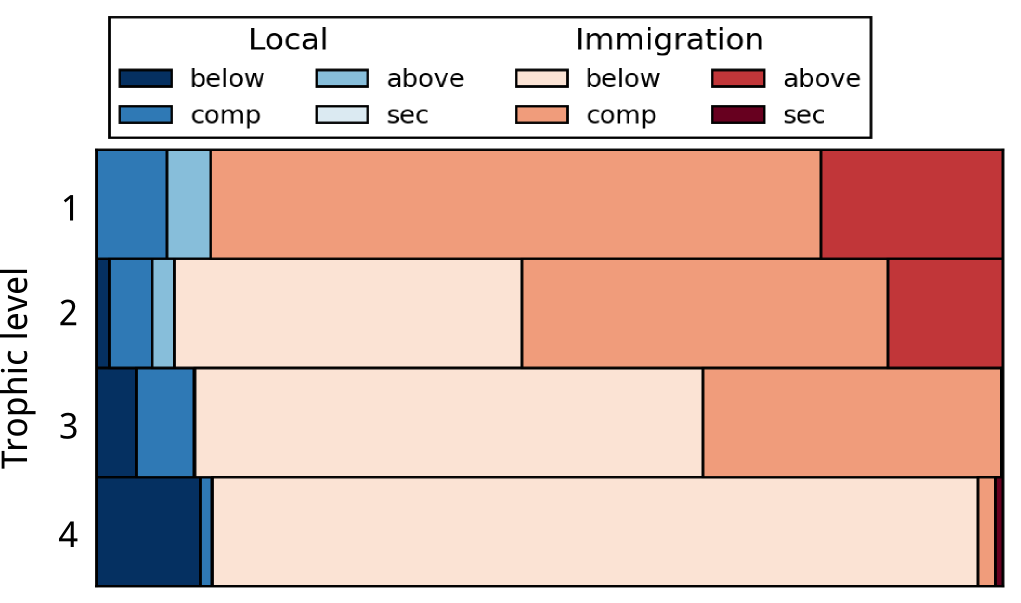}
    \caption{Extinction causes for species on different trophic levels, again for a system of 400 habitats and a dispersal rate of 10. Blue shades refer to local processes, i.e., to speciation events, red shades refer to processes triggered by immigrants from neighbouring habitats.
    We sort events into classes depending on whether the  change occurs in the same trophic level (competition) or in the trophic level below or above.
    Basal species are mostly affected by incoming competitors, whilst higher trophic species are more sensitive to all kind of changes in the network, e.g.~a rearrangement in the lower trophic level.
    }
    \label{fig:reasons for extinction}
\end{figure}

What distinguishes basal species from higher trophic species in our model is food availability.
Basal species incur no risks related to their food source - the resource is present in the same amount in all habitats.
This means that basal species are mainly regulated by competition.
Fig. \ref{fig:reasons for extinction} illustrates this.
It shows the extent of various causes of extinction for the different trophic levels.
Basal species mainly go extinct because of local emergence of a competitor or the incoming of a competitor from neighbouring habitats. Higher-level species have to overcome another obstacle. 
They need to find a suitable prey, which is not the same in all habitats.
So predators depend strongly on their trophic environment.
The higher the trophic position of a species the less prey species are present.
This enhances the influence of the biotic environment.
A look at the extinction reasons for higher trophic levels in Fig. \ref{fig:reasons for extinction} shows that the chance of going extinct due to a non-competitive reason increases with trophic level.

\section{Discussion}
We have introduced and investigated a spatially explicit evolutionary food web model that allows us to explore the distribution of species in space and time, as well as the waxing and waning of species ranges with time. This is the first model that makes it possible to explore these features in the context of trophic networks, showing how they are influenced by competition as well as by predators and prey. 
Indeed, our model produces empirically well-known patterns in space and time, such as lifetime distributions, species-area relationships, distance decay of similarity, and temporal change of geographic range. On top, we obtain a variety of additional result and obtain insights into the mechanisms that generate these patterns. While one might argue that some patterns need to trivially emerge in a model like ours, the multitude of patterns emerging together is remarkable.

We find that most species only appear for short times and have small ranges.
For species that conquer larger portions of the web we analysed the shape of the range evolution and found that basal species show the empirically observed ``hat'' pattern more often than species in higher trophic levels.
This indicates that the trophic position of a species plays a major role for its range expansion success as well as the shape of its range expansion trajectory over time.
To our knowledge, this has not been discussed in the literature so far.

Recently Zliobaite et al. \cite{Zliobaite2017a} analyzed which factors are more correlated to the rise and fall of the range expansion trajectory in fossil data sets of basal mammals. 
The range expansion follows the so called ``hat pattern'' that consists of five phases in a species lifetime: Origination, expansion, peak, decline and extinction.
They found that the temporal location of the peak of the hat pattern is more impacted by competition while the brims are more influenced by abiotic environmental factors.
They also compared the range curves of different random walk models with the shape of empirical data and found that a random walk model with competition and environmental factor provides the most realistic looking curves.

Our results for basal species  provide an illustration of the mechanism that might  lead to these findings.
There is one major difference in assumptions: The ``environment'' in our case is the trophic environment (network structure and abundance distribution).
We do not model an abiotic surrounding, yet basal range curves look strikingly hat shaped.
A species needs to fit into this trophic environment (network) to first establish a viable population (origination).
To successfully increase its range it needs to disperse to neighbouring habitats and be a viable competitor there as well.
This leads to the extinction of another species as the dispersing competitor takes its place.
As neighbouring basal communities are similar in our systems, the chances are high that the species can spread on a large portion on the grid replacing other species (expansion).
This continues until the species has reached the maximum range (peak).
It is only a matter of time then until this process is repeated with the species having become the inferior competitor.
The species is then successively replaced by a better adapted species (decline).
The species thus  ages as the network structure changes.
At some point the species has vanished on all habitats (extinction).
This means that we observe the same dynamics as suggested by Zliobaite et al., but with the trophic and not the abiotic environment as the main driver of the initial increase and later decrease of the range.
The truth is probably that both the trophic and abiotic environment are important, as both play an important role in real ecosystems.

Regarding higher trophic species in our system the range expansion curves look more diverse and often do not resemble the hat shape.
These species depend on the composition on the layer below.
As this layer changes the fate of the higher species changes as well.
The emergence of a new prey species that spreads over the network can save a consumer species from extinction.
This cannot happen for basal species as these are more or less completely controlled by competition.
The studies that we know often deal with basal species, so we are not convinced that the hat pattern is ubiquitous for all species.
Future empirical work could focus on predator range expansion and try to find a case where a predator species could regain its range after the emergence of a new prey.

A qualitative difference between the basic trophic layer and higher layers occurred in our data also with respect to the similarity of networks in nearby habitats. The similarity index decays particularly slowly for the basal layer.
Theory on distance decay suggests that spatial heterogeneity is a main driver in community turnover in two ways: (1) Competitive species sorting along environmental gradients and (2) topological influences that let species with different dispersal abilities experience different landscapes \cite{Nekola1999}.
As we use a homogeneous landscape we expect the first driver to be non-existent for the basal layer, as all habitats hold the same type and amount of resource.
As species do not fundamentally differ in their dispersal abilities and we do not have a heterogeneous spatial topology, the second point is only weakly relevant for the basal species.
They have slightly different chances of being chosen for dispersal as we choose the next coloniser depending on the biomass density. As we have seen, biomass densities are quite similar for basal species.
What remains is a temporal aspect: species that are older can reside on more habitats and have thus a higher chance of being chosen to disperse.
To put the cart before the horse, this confirms the theory on distance decay: 
We expect a much faster decay in a heterogeneous environment, and this is exactly what we observe for higher trophic layers, which experience heterogeneity due to the spatial turnover in basal species composition.
A trend to faster decay rates in higher trophic levels was also found in a meta-study\cite{Soininen2007}.

A model is always a simplification of reality.
Some of the  assumptions underlying our model are worth discussing.
Species in nature are not restricted to the one-dimensional niche space that we assume.
In fact, the original Webworld model characterised species by a large vector of traits \cite{Caldarelli1998}.
We, in contrast, characterized species by three traits, all of which are based on body mass. We think that this is the reason why  we do not observe super abundant species, but species densities are all of a similar order of magnitude. 
With a higher-dimensional trait space there must exist more diverse species types and probably also super abundant species, which  have a globally optimal trait vector.
Nevertheless, our simplification leads to an overall shape of the rank abundance curves that is realistic, as one would expect from a niche apportionment model\cite{Tokeshi1990}.
Harpole and Tilman showed that diversity and evenness of grassland communities decreased when niche dimensionality was reduced by adding limiting nutrients to plot experiments \cite{Harpole2007}.
In turn this indicates that rank abundance curves for less dimensional communities will be flatter.
This is in line with the shape of our rank abundance curves. 

Our choice of parameters is guided by the aim to make the model feasible. 
To be able to perform computer simulations on a large number of habitats we chose a relatively small value for the amount of resource $R$, so that the number of species of a local food web remained below 25.
As we wanted to simulate food webs and not only basal communities we needed to choose a value of the efficiency $\lambda$ that allows for the emergence of several trophic layers. 
In the original Webworld Model, $\lambda$ was identified with the proportion of biomass that is passed from one trophic layer to the next. 
The value of 0.65 that we use here is much larger than the empirically established value of 0.1\cite{Lindemann1942}.
However, in the original Webworld model no distinction was made between resident biomass and biomass fluxes. Therefore the variable $B$ was identified with biomass, while its occurrence in equation \eqref{eq: population} in fact suggests that it represents biomass flux. 
A further reason for the difference between the model value and the empirical value of $\lambda$ is that the model does not take into account energy input due to the below-ground ecological processes. Due to all these simplifications of the model, we do not consider it important to provide an empirical justification of the precise value of the parameter $\lambda$, but base its value on the condition that the model yields food webs with several trophic levels. 

In contrast to other models, the model used here does not rely on an extrinsic extinction rate that randomly extirpates species that might be well adapted to the network.
All extinction events are driven by the trophic dynamics, yet we observe an ongoing species turn over.
We thus study the pure food web dynamics without a heterogeneous or fluctuating environment and still observe ecological reasonable species distributions.
This indicates that incorporating abiotic environments and their fluctuations is not necessarily needed to study food web dynamics.

Rogge et al.~analysed lifetime distributions and SAR curves in a model that is simpler than ours as it does not include population sizes \cite{Rogge2019}.
The lifetime distributions that we find are considerably steeper (slope $-2.4$) compared to their value around $-1.7$.
It is also larger than the values reported for empirical findings, which lie between 1 and 2; Newman and Palmer pin them down to $1.7\pm0.3$ \cite{Newman1999}. Data for contemporary lifetime distributions show a power-law like shape\cite{Keitt1998, Bertuzzo2011} with exponents that are in agreement with the exponent of paleological data\cite{Newman1999}.
It is noteworthy that there is no consensus whether lifetime distributions follow a power-law or an exponential law, as data often allow for both types of fit, due to (large) uncertainties in fossil data \cite{Newman1997, Newman1999}. Exponentials, of course, have a changing slope in a double-logarithmic plot and can thus also be compatible with the exponent observed by us.

Curiously, our model shows steep lifetime distributions even though there is no external random extinction implemented as in other models\cite{Rogge2019}.
One implication of our value $\alpha > 2$ is that our distribution has a well-defined mean. This features is shared with exponential distributions.

McPeek argues that lifetime distributions depend on the number and survival time of ``transient'' species, i.e. species that are on their way to extinction\cite{McPeek2007}. He reasons that the time to extinction is elongated for species that are similar, because the inferior competitor holds out longer when it competes with more similar species.
If this applies to our type of model, this indicates that species in our system are, despite the one-dimensional niche axis, not as similar as species in the model of Rogge et al.\cite{Rogge2019} that uses the same niche axis, as we observe shorter lifetimes. The difference is that interaction links in\cite{Rogge2019} are binary (presence versus absence), whilst we use Gaussian feeding kernels.  The fact that this difference affects the lifetime distributions emphasises the importance of considering details of the trophic interactions.
The SAR curves on the contrary are flatter in our model than in the model by Rogge et al.~ and in better agreement with empirical data.

O'Sullivan et al. \cite{OSullivan2019} found in a competitive metacommunity assembly models a similar collection of macroecological patterns (SAD, range size distribution RSD and SAR) as we did, when regional diversity was near equilibrium.
They refer to the work of McGill\cite{McGill2010} who analysed the assumptions underlying models of macroecological patterns and found that three key ingredients seem to be sufficient for such  patterns to emerge.
Those are a left skewed SAD, clumping of populations in space, and species distributions in space that are uncorrelated from other species spatial distributions.
O'Sullivan et al.\cite{OSullivan2019} report that all three ingredients occur in their model and are shaped by regional diversity equilibrium.
The closer the system to regional equilibrium the stronger are the observed key patterns (SAD, RSD, Spatial non-correlation).
They relate their finding with the theory of ecological structural stability, which revolves around the dynamics on a regional scale.
Our communities, in contrast, are trophic communities, operate always near local and regional species equilibrium, i.e. in the regime where O'Sullivan and coauthors\cite{OSullivan2019} find the most prominent form of the basic patterns.
Comparing the patterns we observe, we also see SADs that are left skewed, and a local clumping of species.
We did not analyse the spatial correlation between species.
As we have trophic layers of species there will be some correlation between predators and their prey as they can only persist in a habitat if prey is present.
In addition to the results obtained by  O'Sullivan et al.~\cite{OSullivan2019},   we also derive liefetime distributions, i.e., a paleoecological pattern that also seem to be connected to the metacommunity dynamics.
This might indicate that spatial non-correlation is not the most important factor in the mechanisms producing macroecological patterns.

To conclude, our evolutionary food web model produces empirically well studied ecological and paleological patterns. We thus are armed with a valuable tool to broaden our understanding of the mechanisms behind those patterns. Our findings that trophic position influences geographic range and lifetime of a species  might motivate further work regarding the interplay of abiotic and trophic factors on range expansion on evolutionary time scales. 

More generally, evolutionary models can assist us in forming a deeper knowledge of the processes that lead to what is remnant in fossils. As recently pointed out by Marshall \cite{Marshall2017} in his fifth law of paleobiology, extinction erases information. It is a strength of evolutionary food web models that they allow us to study processes whose extent eludes direct observations.

\bibliographystyle{unsrt} 
\bibliography{Bibliography.bib}

\begin{thebibliography}{10}

\bibitem{Levin1992}
Simon~A. Levin.
\newblock The problem of pattern and scale in ecology: The robert h. macarthur
  award lecture.
\newblock {\em Ecology}, 73(6):1943--1967, 12 1992.

\bibitem{McGill2007}
Brian~J. McGill, Rampal~S. Etienne, John~S. Gray, David Alonso, Marti~J.
  Anderson, Habtamu~Kassa Benecha, Maria Dornelas, Brian~J. Enquist, Jessica~L.
  Green, Fangliang He, Allen~H. Hurlbert, Anne~E. Magurran, Pablo~A. Marquet,
  Brian~A. Maurer, Annette Ostling, Candan~U. Soykan, Karl~I. Ugland, and
  Ethan~P. White.
\newblock Species abundance distributions: moving beyond single prediction
  theories to integration within an ecological framework.
\newblock {\em Ecology Letters}, 10(10):995--1015, 2007.

\bibitem{Brown1996}
James~H. Brown, George~C. Stevens, and Dawn~M. Kaufman.
\newblock The geographic range: Size, shape, boundaries, and internal
  structure.
\newblock {\em Annual Review of Ecology and Systematics}, 27(1):597--623, 1996.

\bibitem{Gaston1996}
Kevin~J. Gaston.
\newblock Species-range-size distributions: patterns, mechanisms and
  implications.
\newblock {\em Trends in Ecology \& Evolution}, 11(5):197 -- 201, 1996.

\bibitem{Gaston2003}
Kevin~J. Gaston.
\newblock {\em The structure and dynamics of geographic ranges}.
\newblock Oxford University Press, 2003.

\bibitem{Nekola1999}
Jeffrey~C. Nekola and Peter~S. White.
\newblock The distance decay of similarity in biogeography and ecology.
\newblock {\em Journal of Biogeography}, 26(4):867--878, 1999.

\bibitem{Soininen2007}
Janne Soininen, Robert McDonald, and Helmut Hillebrand.
\newblock The distance decay of similarity in ecological communities.
\newblock {\em Ecography}, 2007.

\bibitem{Tjoerve2003}
Even Tjørve.
\newblock Shapes and functions of species–area curves: a review of possible
  models.
\newblock {\em Journal of Biogeography}, 30(6):827--835, 2003.

\bibitem{Drakare2006}
Stina Drakare, Jack~J. Lennon, and Helmut Hillebrand.
\newblock The imprint of the geographical, evolutionary and ecological context
  on species–area relationships.
\newblock {\em Ecology Letters}, 9(2):215--227, 2006.

\bibitem{Newman1999}
M.~E.~J. Newman and R.~G. Palmer.
\newblock Models of extinction.
\newblock {\em arXiv}, 1999.

\bibitem{Zliobaite2017a}
Indre Zliobaite, Mikael Fortelius, and Nils~C. Stenseth.
\newblock Reconciling taxon senescence with the red queen’s hypothesis.
\newblock {\em Nature}, 552, 2017.

\bibitem{Liow2007}
Lee~Hsiang Liow and Nils~Chr. Stenseth.
\newblock The rise and fall of species: implications for macroevolutionary and
  macroecological studies.
\newblock {\em Proceedings of the Royal Society B}, 2007.

\bibitem{Foote2007}
Michael Foote, James~S. Crampton, Alan~G. Beu, Bruce~A. Marshall, Roger~A.
  Cooper, Phillip~A. Maxwell, and Iain Matcham.
\newblock Rise and fall of species occupancy in cenozoic fossil mollusks.
\newblock {\em Science}, 318, 2007.

\bibitem{Valen1973}
Leigh~Van Valen.
\newblock A new evolutionary law.
\newblock {\em Evolutionary Theory}, 1973.

\bibitem{Finnegan2008}
Seth Finnegan, Jonathan L-Payne, and Steve~C. Wang.
\newblock The red queen revisited: reevaluating the age selectivity of
  phanerozoic marine genus extinctions.
\newblock {\em Paleobiology}, 2008.

\bibitem{McGill2010}
Brian~J. McGill.
\newblock Towards a unification of unified theories of biodiversity.
\newblock {\em Ecology Letters}, 13(5):627--642, 2010.

\bibitem{Rosenzweig1995}
Michael~L. Rosenzweig.
\newblock {\em Species diversity in space and time}.
\newblock Cambridge University Press, 1995.

\bibitem{OSullivan2019}
Jacob~D. O'Sullivan, Robert~J. Knell, and Axel~G. Rossberg.
\newblock Metacommunity-scale biodiversity regulation and the self-organised
  emergence of macroecological patterns.
\newblock {\em Ecology Letters}, 2019.

\bibitem{May2016}
Felix May, Thorsten Wiegand, Sebastian Lehmann, and Andreas Huth.
\newblock Do abundance distributions and species aggregation correctly predict
  macroecological biodiversity patterns in tropical forests?
\newblock {\em Global Ecology and Biogeography}, 2016.

\bibitem{Willis1922}
J.~C. Willis.
\newblock {\em Age and area; a study in geographical distribution and origin of
  species}.
\newblock Cambrindge [Eng.] The University Press, 1922.

\bibitem{Gaston2000}
Kevin~J. Gaston and Tim~M. Blackburn.
\newblock {\em Pattern and Process in Macroecology}.
\newblock Blackwell Science, 2000.

\bibitem{Rogge2019}
Tobias Rogge, David Jones, Barbara Drossel, and Korinna~T. Allhoff.
\newblock Interplay of spatial dynamics and local adaptation shapes species
  lifetime distributions and species--area relationships.
\newblock {\em Theoretical Ecology}, Feb 2019.

\bibitem{Bertuzzo2011}
Enrico Bertuzzo, Samir Suweis, Lorenzo Mari, Amos Maritan, Ignacio
  Rodr{\'\i}guez-Iturbe, and Andrea Rinaldo.
\newblock Spatial effects on species persistence and implications for
  biodiversity.
\newblock {\em Proceedings of the National Academy of Sciences},
  108(11):4346--4351, 2011.

\bibitem{Dunne2014}
Jennifer~A. Dunne, Conrad~C. Labandeira, and Richard~J. Williams.
\newblock Highly resolved early eocene food webs show development of modern
  trophic structure after the end-cretaceous extinction.
\newblock {\em Proceedings of the Royal Society B: Biological Sciences},
  281(1782):20133280, 2014.

\bibitem{McPeek2007}
Mark~A McPeek.
\newblock The macroevolutionary consequences of ecological differences among
  species.
\newblock {\em Paleontology}, 2007.

\bibitem{Durrett1996}
Rick Durrett and Simon Levin.
\newblock Spatial models for species-area curves.
\newblock {\em Journal of Theoretical Biology}, 179(2):119 -- 127, 1996.

\bibitem{Rosindell2007}
James Rosindell and Stephen~J. Cornell.
\newblock Species-area relationships from a spatial explicit neutral model in
  an infinite landscape.
\newblock {\em Ecology Letters}, 2007.

\bibitem{Allhoff2015a}
Korinna~T. Allhoff, Eva~Marie Weiel, Tobias Rogge, and Barbara Drossel.
\newblock On the interplay of speciation and dispersal: An evolutionary food
  web model in space.
\newblock {\em Journal of Theoretical Biology}, 366:46 -- 56, 2015.

\bibitem{Loeuille2008}
N.~Loeuille and M.A. Leibold.
\newblock Ecological consequences of evolution in plant defenses in a
  metacommunity.
\newblock {\em Theoretical Population Biology}, 74(1):34 -- 45, 2008.

\bibitem{Bolchoun2017}
Lev Bolchoun, Barbara Drossel, and Korinna~T. Allhoff.
\newblock Spatial topologies affect local food web structure and diversity in
  evolutionary metacommunities.
\newblock {\em Scientific Reports}, 7, 2017.

\bibitem{Caldarelli1998}
Guido Caldarelli, Paul~G. Higgs, and Alan~J. McKane.
\newblock Modelling coevolution in multispecies communities.
\newblock {\em Journal of Theoretical Biology}, 193(2):345 -- 358, 1998.

\bibitem{Loeuille2005}
Nicolas Loeuille and Michel Loreau.
\newblock Evolutionary emergence of size-structured food webs.
\newblock {\em Proceedings of the National Academy of Sciences},
  102(16):5761--5766, 2005.

\bibitem{Allhoff2013}
Korinna~T. Allhoff and Barbara Drossel.
\newblock When do evolutionary food web models generate complex networks?
\newblock {\em Journal of Theoretical Biology}, 334:122 -- 129, 2013.

\bibitem{Allhoff2015}
KT~Allhoff, D~Ritterskamp, BC~Rall, B~Drossel, and C~Guill.
\newblock Evolutionary food web model based on body masses gives realistic
  networks with permanent species turnover.
\newblock {\em Scientific Reports}, 5, 2015.

\bibitem{Brown2004}
James~H. Brown, James~F. Gillooly, Andrew~P. Allen, Van~M. Savage, and
  Geoffrey~B. West.
\newblock Toward a metabolic theory of ecology.
\newblock {\em Ecology}, 85(7):1771--1789, 2004.

\bibitem{Hirt2017}
Myrian~R. Hirt, Walter Jetz, Björn~C. Rall, and Ulrich Brose.
\newblock A general scaling law reveals why the largest animals are not the
  fastest.
\newblock {\em Nature Ecology \& Evolution}, 1:1116--1122, 2017.

\bibitem{Brose2010}
Ulrich Brose.
\newblock Body-mass constraints on foraging behaviour determine population and
  food-web dynamics.
\newblock {\em Functional Ecology}, 24(1):28--34, 2010.

\bibitem{Tokeshi1990}
Mutsunori Tokeshi.
\newblock Niche apportionment or random assortment: species abundance patterns
  revisited.
\newblock {\em Journal of Animal Ecology}, 59(3):1129--1146, 1 1990.

\bibitem{Harpole2007}
W.~Stanley Harpole and David Tilman.
\newblock Grassland species loss resulting from reduced niche dimension.
\newblock {\em Nature}, 2007.

\bibitem{Lindemann1942}
Raymond Lindemann.
\newblock The trophic-dynanic aspect of ecology.
\newblock {\em Ecology}, 1942.

\bibitem{Keitt1998}
Toimothy~H. Keitt and H.~Eugene Stanley.
\newblock Dynamics of north american breeding bird populations.
\newblock {\em Nature}, 1998.

\bibitem{Newman1997}
M.E.J. Newman.
\newblock A model of mass extinction.
\newblock {\em Journal of Theoretical Biology}, 189(3):235 -- 252, 1997.

\bibitem{Marshall2017}
Charles~R. Marshall.
\newblock Five paleobiological laes needed to understand the evolution of the
  living biota.
\newblock {\em Nature Ecology \& Evolution}, 1, 2017.

\end{thebibliography}

\section*{Acknowledgements}
This work was supported by the German Research Foundation (DFG) under contract number Dr300/12-2. We acknowledge support by the Open Access Publishing fund of the Technische Universit\"at Darmstadt.

\newpage

\end{document}